\documentclass[12pt]{article}
\usepackage{amsfonts}
\usepackage{amsmath}
\usepackage{cite}

\csname @addtoreset\endcsname{equation}{section}
\newcommand{\eq}{\begin{equation}}
\newcommand{\feq}{\end{equation}}
\newcommand{\eqn}{\begin{eqnarray}}
\newcommand{\feqn}{\end{eqnarray}}
\newcommand{\arr}{\begin{eqnarray*}}
\newcommand{\farr}{\end{eqnarray*}}

\font\mybb=msbm10 at 12pt
\def\bb#1{\hbox{\mybb#1}}

\def\bR {\bb{R}}

\begin{document}

\begin{titlepage}
\begin{flushright}
IFUM-732-FT\\
hep-th/0212096
\end{flushright}
\vspace{.3cm}
\begin{center}
\renewcommand{\thefootnote}{\fnsymbol{footnote}}
{\Large \bf AdS$_2$ Supergravity and Superconformal Quantum Mechanics}
\vskip 25mm
{\large \bf {Marco Astorino$^1$\footnote{marcoa@pcteolaur1.mi.infn.it},
Sergio Cacciatori$^{2,3}$\footnote{cacciatori@mi.infn.it},
Dietmar Klemm$^{1,3}$\footnote{dietmar.klemm@mi.infn.it}
and Daniela Zanon$^{1,3}$\footnote{daniela.zanon@mi.infn.it}}}\\
\renewcommand{\thefootnote}{\arabic{footnote}}
\setcounter{footnote}{0}
\vskip 10mm
{\small
$^1$ Dipartimento di Fisica dell'Universit\`a di Milano,\\
Via Celoria 16, I-20133 Milano.\\

\vspace*{0.5cm}

$^2$ Dipartimento di Matematica dell'Universit\`a di Milano,\\
Via Saldini 50, I-20133 Milano.\\

\vspace*{0.5cm}

$^3$ INFN, Sezione di Milano,\\
Via Celoria 16,
I-20133 Milano.\\
}
\end{center}
\vspace{2cm}
\begin{center}
{\bf Abstract}
\end{center}
{\small We investigate the asymptotic dynamics of topological anti-de~Sitter
supergravity in two dimensions. Starting from the formulation as a BF theory,
it is shown that the AdS$_2$ boundary conditions imply that the asymptotic symmetries
form a super-Virasoro algebra. Using the central charge of this algebra in
Cardy's formula, we exactly reproduce the thermodynamical entropy of AdS$_2$
black holes. Furthermore, we show that the dynamics of the dilaton and its
superpartner reduces to that of superconformal transformations that leave
invariant one chiral component of the stress tensor supercurrent of a two-dimensional
conformal field theory. This dynamics is governed by a supersymmetric extension
of the de~Alfaro-Fubini-Furlan model of conformal quantum mechanics.
Finally, two-dimensional de~Sitter gravity is also considered, and the dS$_2$ entropy
is computed by counting CFT states.
}

\end{titlepage}

\section{Introduction}

\label{intro}

There is by now a great deal of evidence that the correspondence
between type IIB string theory on AdS$_5$ $\times$ S$^5$ and
${\cal N}=4$ super-Yang-Mills theory in four
dimensions \cite{Maldacena:1997re} extends
to a more general duality between any gravitational theory on
anti-de~Sitter spaces and conformal field theories residing on the
boundary of AdS \cite{Witten:1998qj}. Such a duality represents
an explicit realization of the holographic principle \cite{'tHooft:gx}.
In three dimensions, pure Einstein-AdS gravity was known to be
related to a two-dimensional conformal field theory \cite{Coussaert:1995zp}
even before the advent of the AdS/CFT correspondence. This is based
on the fact that pure Einstein gravity in three dimensions can be written
as a Chern-Simons theory \cite{Achucarro:1986vz,Witten:1988hc}, which is known to
reduce to a WZNW model in presence of a boundary \cite{Elitzur:1989nr}.
The boundary conditions for asymptotically AdS spaces \cite{Brown:nw}
provide then the constraints for a Hamiltonian reduction from
the WZNW model to Liouville field
theory \cite{Coussaert:1995zp}\footnote{The results of \cite{Coussaert:1995zp}
have been generalized to three-dimensional extended AdS supergravity
in \cite{Henneaux:1999ib}.}.
Recently the equivalence of pure gravity in three dimensions
and two-dimensional Liouville theory has been extended
in \cite{Cacciatori:2001un,Klemm:2002ir} to the case of positive cosmological
constant, providing thus an explicit example of the proposed correspondence
between de~Sitter gravity and Euclidean conformal field
theories \cite{Hull:1998vg,Strominger:2001pn,Klemm:2001ea}\footnote{Cf.~\cite{Park:1998qk}
for earlier work.}.

Much less is known for two-dimensional anti-de~Sitter
gravity \cite{Strominger:1998yg}, which, in the spirit of AdS/CFT,
should have a dual description in terms of a conformal quantum
mechanical system \cite{Claus:1998ts}\footnote{Cf.~also
\cite{Cacciatori:1999rp} and references therein.}. Progress in this
direction has been made in \cite{Cadoni:2000gm}, where it was shown
that two-dimensional Jackiw-Teitelboim-AdS gravity \cite{Teitelboim:ux}
induces on the spacetime boundary a conformally invariant dynamics that
can be described in terms of a de~Alfaro-Fubini-Furlan (DFF)
model \cite{deAlfaro:1976je} coupled to an external
source\footnote{Cf.~\cite{Brigante:2002rv} for related work.}.

In the present paper we will extend the results
of \cite{Cadoni:2000gm,Brigante:2002rv} to topological AdS$_2$
supergravity. Making use of its gauge theory
formulation \cite{Isler:1989hq,Chamseddine:1990wn},
we will be able to show in a clear and transparent way how
the superconformal algebra appears in this system, and how the dynamics
can be reduced to that of a superconformally invariant quantum mechanics.

The remainder of this paper is organized as follows: In section
\ref{topgaugeth}, the formulation of topological AdS$_2$ supergravity as
a gauge theory (BF theory) is briefly reviewed. In section
\ref{assymm}, we translate the boundary conditions for asymptotically
AdS$_2$ spaces \cite{Cadoni:1998sg,Cadoni:1999ja} into conditions
for the bosonic fields appearing in the gauge theory. Furthermore,
suitable boundary conditions on the fermions are defined.
Subsequently, we show that the gauge transformations preserving this
asymptotic behaviour generate a super-Virasoro algebra, and compute its
central charge. Using this in Cardy's formula for the
asymptotic level density, we exactly reproduce the thermodynamical entropy
of AdS$_2$ black holes.
In the following section, it is shown that the dynamics of the dilaton and its
superpartner reduces to that of superconformal transformations that leave
invariant one chiral component of the stress tensor supercurrent of a two-dimensional
conformal field theory. This dynamics turns out to be governed by a
supersymmetric extension of the DFF model of conformal
quantum mechanics.
In section \ref{dS},
we generalize the considerations of sections \ref{assymm}-\ref{redsupconf},
which refer to AdS gravity, to the de~Sitter case.
Again, a microstate counting yields precisely the entropy of two-dimensional
de~Sitter space.
We close our paper with some final remarks, followed by an appendix that
contains our conventions.

\section{Two-dimensional anti-de~Sitter supergravity as topological gauge theory}
\label{topgaugeth}

We consider the minimal supersymmetric extension of the
Jackiw-Teitelboim (JT) model \cite{Teitelboim:ux}, which contains the zweibein
$e^a_{\mu}$, a Rarita-Schwinger field $\psi_{i\mu}$, the dilaton $\eta$ and
dilatino $\lambda_i$. The action is given by \cite{Chamseddine:1991fg}

\begin{equation}
I = \frac{\beta}{2} \int d^2x \sqrt{-g} \eta (R - 2\Lambda) + \frac{\beta}{4l}\int
    \eta \psi \wedge \gamma_5 \bar{\psi} + \frac{\beta}{4}\int \bar{\lambda} D\psi\,,
    \label{actiongrav}
\end{equation}

where $\beta$ denotes a dimensionless coupling constant, $\Lambda = 1/l^2$ is
the cosmological constant\footnote{Notice that for the $(+,-)$ signature used
here, $\Lambda$ is positive for AdS and negative for dS.} and

\begin{equation}
(D\psi)_j = d\psi_j - \frac 14 (\gamma_{ab})^k_{\;\; j} \omega^{ab}\wedge \psi_k +
            \frac{i}{2l}(\gamma_a)^k_{\;\; j} e^a \wedge \psi_k\,.
\end{equation}

$\psi_{i\mu}$ and $\lambda_i$ are Majorana spinors, i.~e.
$\lambda_i = \epsilon_{ij}\bar{\lambda}^i$ etc., where $\epsilon_{12} = 1$.

The scalar curvature $R$ is calculated using the spin connection that solves the
torsion-free condition

\begin{equation}
T^a = d e^a + \omega^a_{\;\; b} \wedge e^b + \frac i4 (\gamma^a)^j_{\;\; k} \psi^k \wedge
\psi_j = 0\,. \label{torsion}
\end{equation}

Notice that, throughout the paper,
the product of two fermions differs by a factor $i$ from the standard Grassmann
product fulfilling $(ab)^{\ast} = b^{\ast} a^{\ast}$.

The supergravity model (\ref{actiongrav}) can be formulated as an
Osp$(1|2)$ topological gauge theory \cite{Chamseddine:1990wn}.
This is similar to the three-dimensional case,
where the Einstein-AdS supergravity action can be written as a sum
of two Chern-Simons actions \cite{Achucarro:1986vz}.
Let us introduce the superconnection

\begin{equation}
\Gamma = A^A \tau_A + \psi_i R^i \equiv A + \psi\,, \label{Gamma}
\end{equation}

and the scalar multiplet

\begin{equation}
S = \phi^A \tau_A + \lambda_i R^i \equiv \phi + \lambda \label{S}
\end{equation}

in the adjoint representation of Osp$(1|2)$. In (\ref{Gamma}) and
(\ref{S}), the $\tau_A$ and $R^i$ are respectively the bosonic and
fermionic generators of the graded Lie algebra osp$(1|2)$ (cf.~appendix
\ref{convAdS}).

The action describing AdS$_2$ supergravity is given
by\footnote{(\ref{actiongauge}) can also be obtained
from the super-Chern-Simons action
$I = \frac{k}{4\pi}\int {\mathrm{STr}}(\hat{\Gamma} \wedge
d\hat{\Gamma} + \frac 23 \hat{\Gamma} \wedge \hat{\Gamma} \wedge \hat{\Gamma})$
using the dimensional reduction ansatz $\hat{\Gamma} = \Gamma + S dz$.}

\begin{equation}
I = \frac{\beta}{2}\int {\mathrm{STr}}(S{\cal F})\,, \label{actiongauge}
\end{equation}

where

\begin{equation}
{\cal F} = d\Gamma + \Gamma \wedge \Gamma \label{fieldstr}
\end{equation}

denotes the Osp$(1|2)$ field strength.

The action (\ref{actiongauge}) is invariant under the gauge
transformations

\begin{eqnarray}
\delta_{\Xi} \Gamma &=& d\Xi + [\Gamma, \Xi]\,, \label{gaugetransfGam} \\
\delta_{\Xi} S &=& [S, \Xi]\,. \label{gaugetransfS}
\end{eqnarray}

The equations of motion following from (\ref{actiongauge}) read

\begin{eqnarray}
{\cal F} &=& 0\,, \label{F=0} \\
DS &\equiv& dS + [\Gamma, S] = 0\,, \label{DS=0}
\end{eqnarray}

so that the solutions are those of flat Osp$(1|2)$ connections in
two dimensions.\\
To make contact with gravity,
one decomposes the fields $A$ and $\phi$ according to

\begin{equation}
A^A = (l^{-1} e^a, \omega)\,, \qquad \phi^A = (\phi^a, 4\eta)\,,
      \label{decomp}
\end{equation}

where $\omega = \omega^0_{\;\; 1}$ and $a = 0,1$.
Using (\ref{decomp}), the bosonic part $F \equiv dA + A \wedge A +
\psi \wedge \psi$ of the field strength (\ref{fieldstr}) reads

\begin{equation}
F^{A} = (l^{-1} T^a, d\omega + l^{-2} e^0 \wedge e^1 - \frac 14 \psi \wedge
         \gamma_5 \bar{\psi})\,, \label{fieldstrdec}
\end{equation}

where $T^a$ will correspond to the torsion tensor given in (\ref{torsion}).
Inserting (\ref{fieldstrdec}) into the action (\ref{actiongauge}),
we see that vanishing torsion
is enforced by the Lagrange multipliers $\phi^a$.
Eliminating the $\phi^a$ by going to the second order formalism, and
rescaling $\psi \to \psi/\sqrt l$, $\lambda \to \lambda \sqrt l$, we
recover (\ref{actiongrav})\footnote{The formulation of the supergravity
theory defined by the action (\ref{actiongrav}) as a gauge theory
involves the identification of non-abelian gauge transformations with
gravitational symmetries. For subtleties concerning this point
cf.~\cite{Schaller:1993pa}.}.

\section{Asymptotic symmetries}
\label{assymm}

\subsection{Boundary conditions}

An asymptotically AdS$_2$ geometry is one for which the bosonic
fields $A$ and $\phi$ behave for $r \to \infty$ as \cite{Brigante:2002rv}

\begin{eqnarray}
A &=& \left[\frac{r}{l^2} - \frac{T_B(t)}{2\eta_0 r}\right] \tau_0 dt
        + \left[\frac{r}{l^2} + \frac{T_B(t)}{2\eta_0 r}\right] \tau_2 dt
        + \frac{dr}{r} \tau_1\,, \nonumber \\
\phi &=& \left[\alpha(t)\frac{r}{l^2} + \frac{\beta(t)}{r}\right] \tau_0
         + \left[\alpha(t)\frac{r}{l^2} - \frac{\beta(t)}{r}\right] \tau_2
         + \gamma(t) \tau_1\,, \label{asymptA}
\end{eqnarray}

where the functions $T_B(t), \alpha(t), \beta(t), \gamma(t)$,
that depend only on the boundary coordinate $t$, parametrize the subleading
behaviour. $\eta_0$ is a constant to be specified below. As we will see later,
$T_B(t)$ plays the role of the bosonic
part of the energy-momentum tensor of a two-dimensional conformal field theory,
hence the notation.

Actually the asymptotics (\ref{asymptA}) seems slightly more restrictive
than that given in \cite{Brigante:2002rv}, which corresponds to
the boundary conditions defined in \cite{Cadoni:1998sg,Cadoni:1999ja} in
terms of the gravitational fields. However, it will turn out that
(\ref{asymptA}) is still weak enough to allow for an asymptotic symmetry algebra
larger than SL$(2,\bR)$. In other words, (\ref{asymptA}) is, up to trivial
gauge transformations, the most general asymptotically AdS$_2$ solution.
If we translate (\ref{asymptA}) in terms of the metric, we get

\begin{equation}
ds^2 = \left(\frac rl - \frac{T_B(t)l}{2\eta_0 r}\right)^2 dt^2 - \frac{l^2}{r^2} dr^2\,,
       \label{gensol}
\end{equation}

which (for zero fermions) is an {\it exact} solution of the equations of motion
for arbitrary $T_B(t)$\footnote{This is similar to the three-dimensional case,
where two arbitrary functions appear \cite{Banados:1998gg}.}. This means that
(\ref{gensol}) is locally isometric to AdS$_2$; however, solutions with
different $T_B(t)$ are in general not globally equivalent.

Consider now the bosonic background of the supergravity
theory (\ref{actiongrav}) studied in \cite{Christensen:1991dk,Cadoni:1994uf},

\begin{eqnarray}
ds^2 &=& \left(\frac{r^2}{l^2} - a^2\right) dt^2 -
         \left(\frac{r^2}{l^2} - a^2\right)^{-1}dr^2\,, \label{BH} \\
\eta &=& \eta_0 \frac rl\,, \label{eta}
\end{eqnarray}

where $a$ and $\eta_0 \ge 0$ denote integration constants. Following
Ref.~\cite{Cadoni:1994uf}, the solutions with $a^2$ positive, negative or
zero will be denoted by AdS$_2^+$, AdS$_2^-$ and AdS$_2^0$ respectively.
Note that due to the presence of the dilaton, which represents a
position-dependent coupling constant, the solutions with different sign
of $a^2$ are physically inequivalent \cite{Cadoni:1994uf}. In particular,
the solution with positive $a^2$ represents a black hole with event
horizon at $r = al$. The black hole mass and Bekenstein-Hawking
entropy are respectively given by \cite{Cadoni:1994uf}

\begin{equation}
M = \frac{1}{2l}\eta_0 a^2\,, \qquad S = 2\pi \eta_0 a\,.
\label{MS}
\end{equation}

Applying the diffeomorphism $r \mapsto r + a^2l^2/4r$ to the metric (\ref{BH}) yields

\begin{equation}
ds^2 = \left(\frac rl - \frac{a^2l}{4r}\right)^2 dt^2 - \frac{l^2}{r^2} dr^2\,.
\end{equation}

Comparing this with (\ref{gensol}), we see that in this case one has
$T_B(t) = \eta_0 a^2/2 = Ml$, so it essentially coincides with the black hole
mass.

The boundary conditions (\ref{asymptA}) must of course be completed by
giving the asymptotic behaviour of the fermions, which can be obtained
by following the general scheme introduced in \cite{Henneaux:tv,Banados:1998pi}.
Namely, one takes a generic configuration satisfying the above asymptotic
boundary conditions with vanishing fermions, and acts on it with the exact
symmetries of AdS$_2^0$. This will create new terms, from which one can
infer the desired asymptotic behaviour.

In order to determine the exact fermionic symmetries of AdS$_2^0$,
one solves the Killing spinor equation

\begin{equation}
\delta_{\epsilon} \psi = d\epsilon + [A, \epsilon] = 0\,, \label{killspinequ}
\end{equation}

with the connection $A$ given by (\ref{asymptA}) with $T_B(t) = 0$.
The solution $\epsilon = \epsilon_1 R^1 + \epsilon_2 R^2$ of (\ref{killspinequ})
reads

\begin{equation}
\epsilon_{1,2} = \sqrt{\frac lr} \left(1 - \frac{rt}{l^2}\right)
                 \epsilon_0^{(1)} \pm \sqrt{\frac rl} \epsilon_0^{(2)}\,,
                 \label{killspin}
\end{equation}

where $\epsilon_0^{(1,2)}$ denote two arbitrary constant spinors.

Notice that the variation of the dilatino under supersymmetry transformations
generated by $\epsilon$ is given (for a bosonic background with
$\lambda = 0$) by

\begin{equation}
\delta_{\epsilon} \lambda = [\phi, \epsilon]\,. \label{deltalambda}
\end{equation}

The scalar field $\phi$ corresponding to AdS$_2^0$ is

\begin{equation}
\phi = 4\eta_0 \frac rl (\tau_0 + \tau_2)\,, \label{phiAdS0}
\end{equation}

i.~e.~$\alpha(t) = 4\eta_0 l$, $\beta(t) = \gamma(t) = 0$.
Using (\ref{phiAdS0}), one gets

\begin{equation}
[\phi, \epsilon] = 4\eta_0 \sqrt{\frac rl}\left(1 - \frac{rt}{l^2}\right)
                   \epsilon_0^{(1)} (R^1 - R^2)\,,
\end{equation}

so it vanishes only for $\epsilon_0^{(1)} = 0$. This means that the non-zero
scalar field $\phi$ (\ref{phiAdS0}) breaks half of the supersymmetries.
This is the supersymmetric analogue of the observation made
in \cite{Cadoni:2000ah} that a nonvanishing field $\phi$ breaks
the SL$(2,\bR)$ isometry group of AdS$_2$ down to the subgroup of time
translations. We thus encounter a spontaneous partial supersymmetry
breaking: The theory has osp$(1|2)$ invariance, but the vacuum preserves
only an s(2) subalgebra\footnote{s(2) denotes the superalgebra introduced
by Witten to formulate supersymmetric quantum mechanics \cite{Witten:nf}.}.

After this small digression, we come back to the determination of the
boundary conditions for the fermions. Acting on a generic background
(\ref{asymptA}) with the exact AdS$_2^0$ supersymmetries (\ref{killspin}),
one obtains for the gravitino variation (\ref{killspinequ})

\begin{equation}
\delta_{\epsilon}\psi = \frac{T_B(t)}{2\eta_0 \sqrt{rl}}\epsilon_0^{(2)}
                        (R^1 + R^2) dt\,,
\end{equation}

so that we will impose

\begin{equation}
\psi = \frac{T_F(t)}{2\eta_0 \sqrt{2rl}}(R^1 + R^2) dt \label{asymptpsi}
\end{equation}

as the appropriate boundary condition on the gravitinos, where $T_F(t)$ is
an arbitrary function. Below we will see that $T_F(t)$ is the fermionic
superpartner of the bosonic stress tensor $T_B(t)$.

For a generic field $\phi$ given in (\ref{asymptA}),
Eq.~(\ref{deltalambda}) yields a dilatino asymptotics of the form

\begin{equation}
\lambda = \left(\mu(t)\sqrt{\frac{2r}{l}} + \nu(t)\sqrt{\frac{2l}{r}}\right) R^1
          + \left(-\mu(t)\sqrt{\frac{2r}{l}} + \nu(t)\sqrt{\frac{2l}{r}}\right) R^2\,,
          \label{asymptlambda}
\end{equation}

with $\mu(t)$ and $\nu(t)$ some functions.

Notice that the field strength ${\cal F}$ (\ref{fieldstr}) still vanishes
for an arbitrary function $T_F(t)$ in (\ref{asymptpsi}) and for any $T_B(t)$
in (\ref{asymptA}), so that we have
found an exact solution that contains two undetermined functions
$T_{B,F}(t)$. Note also that the functions $\alpha(t)$, $\beta(t)$,
$\gamma(t)$, $\mu(t)$ and $\nu(t)$ that occur in $\phi$ (\ref{asymptA})
and $\lambda$ (\ref{asymptlambda}) are not arbitrary; imposing the
equations of motion (\ref{DS=0}) yields a dynamics for these fields.
In section \ref{redsupconf}, it will be shown that the resulting dynamics is that
of superconformal mechanics.

\subsection{Asymptotic symmetries}

Having now at hand the boundary conditions for both the
bosons (\ref{asymptA}) and fermions (\ref{asymptpsi}), (\ref{asymptlambda}),
we can determine the asymptotic symmetries, i.~e.~the
gauge transformations that preserve these conditions. It will turn
out that the asymptotic symmetries form a super-Virasoro algebra.

Acting on the superconnection $\Gamma$ with an infinitesimal gauge
transformation $\Xi = \Lambda^A \tau_A + \epsilon_i R^i$, we get

\begin{eqnarray}
\delta_{\Xi} \Gamma &=& d\Lambda^A \tau_A + d\epsilon_i R^i \nonumber \\
                    &+& \left(\frac{r}{l^2} + \frac{T_B(t)}{2\eta_0 r}\right)
                        (\Lambda^1 \tau_0 + \Lambda^0 \tau_1 + \frac 12 \epsilon_1 R^1
                        - \frac 12 \epsilon_2 R^2) dt \nonumber \\
                    &+& \left(\frac{r}{l^2} - \frac{T_B(t)}{2\eta_0 r}\right)
                        (-\Lambda^2 \tau_1 + \Lambda^1 \tau_2 + \frac 12 \epsilon_2 R^1
                        - \frac 12 \epsilon_1 R^2) dt \nonumber \\
                    &+& \frac{T_F(t)}{4\sqrt{2rl}\eta_0}[(\epsilon_1 + \epsilon_2)
                        (\tau_2 - \tau_0) - (\epsilon_1 - \epsilon_2)\tau_1 -
                        (\Lambda^0 + \Lambda^2)(R^1 - R^2) - \Lambda^1 (R^1 + R^2)] dt
                        \nonumber \\
                    &+& (-\Lambda^2 \tau_0 - \Lambda^0 \tau_2 + \frac 12 \epsilon_2 R^1
                        + \frac 12 \epsilon_1 R^2)\frac{dr}{r}\,.
\end{eqnarray}

In order to preserve the asymptotic form of the superconnection, the parameters
$\Lambda^A$ and $\epsilon_i$ of the gauge transformation have to be of the form

\begin{eqnarray}
\Lambda^0 &=& \left(\frac{r}{l^2} - \frac{T_B(t)}{2\eta_0 r}\right)\chi(t) +
              \frac{l^2}{2r}\chi''(t) + \frac{l}{4\sqrt 2 \eta_0 r} T_F(t)
              \epsilon_0(t)\,, \nonumber \\
\Lambda^1 &=& -\chi'(t)\,, \nonumber \\
\Lambda^2 &=& \left(\frac{r}{l^2} + \frac{T_B(t)}{2\eta_0 r}\right)\chi(t) -
              \frac{l^2}{2r}\chi''(t) - \frac{l}{4\sqrt 2 \eta_0 r} T_F(t)
              \epsilon_0(t)\,, \nonumber \\
\epsilon_1 &=& \sqrt{\frac{2r}{l}}\epsilon_0(t) + \sqrt{\frac{2l}{r}}
               \left(\frac{T_F(t)}{4 \eta_0 l}\chi(t) - l\epsilon_0'(t)
               \right)\,, \nonumber \\
\epsilon_2 &=& -\sqrt{\frac{2r}{l}}\epsilon_0(t) + \sqrt{\frac{2l}{r}}
               \left(\frac{T_F(t)}{4 \eta_0 l}\chi(t) - l\epsilon_0'(t)
               \right)\,, \label{parameters}
\end{eqnarray}

where $\chi(t)$ and $\epsilon_0(t)$ denote arbitrary bosonic and fermionic
functions respectively. This means that the asymptotic symmetry algebra is in
fact infinite-dimensional.

Under a gauge transformation with the parameters (\ref{parameters}), the fields
$T_B(t)$ and $T_F(t)$ transform as

\begin{eqnarray}
\delta T_B &=& T_B'\chi + 2 T_B\chi' - \frac{c}{12}l^2\chi''' - \frac l2 T_F'
               \epsilon_0 - \frac 32 l T_F\epsilon_0'\,, \label{transfTB} \\
\delta T_F &=& T_F'\chi + \frac 32 T_F\chi' + 2 T_B\epsilon_0 - \frac c3
               l^2 \epsilon_0''\,, \label{transfTF}
\end{eqnarray}

with

\begin{equation}
c = 12\eta_0\,. \label{centralch}
\end{equation}

Eqns.~(\ref{transfTB}) and (\ref{transfTF}) are exactly the transformation laws
of a CFT stress tensor $T_B$ and a spin $3/2$ supercurrent $T_F$ under
superconformal transformations, with a central charge given by
(\ref{centralch})\footnote{By introducing the dimensionless time variable
$\tau = t/l$, and replacing $\chi$ by $l\chi$, (\ref{transfTB}) and
(\ref{transfTF}) assume the usual form, that does not involve the
length scale $l$.}.
This can be easily verified by using the superconformal Ward identities
and the OPE\footnote{Note that the resulting transformation laws are those
of active transformations, whereas (\ref{transfTB}) and (\ref{transfTF}) are
passive, so that some signs have to be changed.}

\begin{eqnarray}
T_B(z) T_B(0) &\sim & \frac{c}{2z^4} + \frac{2}{z^2} T_B(0) + \frac 1z \partial
                      T_B(0)\,, \nonumber \\
T_B(z) T_F(0) &\sim & \frac{3}{2z^2}T_F(0) + \frac 1z \partial T_F(0)\,, \nonumber \\
T_F(z) T_F(0) &\sim & \frac{2c}{3z^3} + \frac 2z T_B(0)\,.
\end{eqnarray}

The transformation rules of the fields $\alpha(t)$, $\beta(t)$, $\gamma(t)$,
$\mu(t)$ and $\nu(t)$ appearing in the scalar $S$, can be readily obtained from
(\ref{gaugetransfS}), but will not be reported here.

Note that the central charge (\ref{centralch}) coincides with the one
computed in \cite{Caldarelli:2000xk,Cadoni:2000gm,Catelani:2000gn}
for the bosonic case by different methods. This means that the central
charge for AdS$_2$ dilaton supergravity is the same as the one for pure
dilaton gravity, similar to the three-dimensional case \cite{Banados:1998pi}.

It is instructive to translate the gauge transformations (\ref{parameters})
in terms of gravitational symmetries. To this end, we consider for a moment only
the bosonic case. On shell, a diffeomorphism $\delta x^{\mu} = v^{\mu}(x)$ is
equivalent to a gauge transformation with parameter
$\Lambda = v^{\mu}A_{\mu}$ \cite{Isler:1989hq}. Thus, the asymptotic symmetries
(\ref{parameters}) correspond to diffeomorphisms

\begin{eqnarray}
\delta t &=& \chi + \frac{l^2 \chi''}{\frac{2r^2}{l^2} - \frac{T_B(t)}{\eta_0}}\,,
             \nonumber \\
\delta r &=& -\chi' r\,,
\end{eqnarray}

that leave the form of the metric (\ref{gensol}) invariant, with
$T_B \to T_B + \delta T_B$, and $\delta T_B$ given by (\ref{transfTB}) with
$\epsilon_0(t) = 0$.

\subsection{Statistical entropy of AdS$_2$ black holes}

In the preceding subsection we found that the asymptotic symmetries
of AdS$_2$ form a superconformal algebra, with central
charge (\ref{centralch}). AdS$_2$ quantum gravity is thus a holomorphic
sector of a superconformal field theory. The asymptotic growth of the
number of states is given by the Cardy formula \cite{Cardy:ie}

\begin{equation}
S = 2\pi\sqrt{\frac{cL_0}{6}}\,. \label{cardy}
\end{equation}

Using $L_0 = Ml$, with the black hole mass $M$ given in (\ref{MS}), and the
central charge (\ref{centralch}), one exactly reproduces the
thermodynamical entropy (\ref{MS}) of the AdS$_2$ black hole.

Similar to the case of three dimensions \cite{Coussaert:jp,Strominger:1997eq},
we are led to identify the zero mass black hole AdS$_2^0$ ($a^2 = 0$ in
(\ref{BH})) with the Ramond ground state of the superconformal field theory,
whereas AdS$_2^-$ with $a^2 = -1$ (which is AdS$_2$ in global coordinates)
corresponds to the Neveu-Schwarz ground state. The NS ground state has
a mass shift

\begin{displaymath}
L_0 = -\frac{c}{24} = -\frac 12 \eta_0\,,
\end{displaymath}

so that $M = L_0/l$ coincides exactly with the mass (\ref{MS}) of AdS$_2^-$
for $a^2 = -1$. Further evidence for this identification comes from the
fact that the covariantly constant spinors for AdS$_2^-$,

\begin{eqnarray*}
\epsilon_1 &=& \left(\epsilon_0^{(1)}\cos\frac{t}{2l} + \epsilon_0^{(2)}
               \sin\frac{t}{2l}\right)\cosh\frac{\zeta}{2} +
               \left(-\epsilon_0^{(1)}\sin\frac{t}{2l} + \epsilon_0^{(2)}
               \cos\frac{t}{2l}\right)\sinh\frac{\zeta}{2}\,, \\
\epsilon_2 &=& -\left(\epsilon_0^{(1)}\cos\frac{t}{2l} + \epsilon_0^{(2)}
               \sin\frac{t}{2l}\right)\sinh\frac{\zeta}{2} +
               \left(\epsilon_0^{(1)}\sin\frac{t}{2l} - \epsilon_0^{(2)}
               \cos\frac{t}{2l}\right)\cosh\frac{\zeta}{2}\,,
\end{eqnarray*}

where $\epsilon_0^{(1,2)}$ denote arbitrary constant spinors,
and $\sinh\zeta \equiv r/l$, are antiperiodic under $t \to t + 2\pi l$.

Notice in this context that the coordinate transformation

\begin{eqnarray}
t &\to & i e^{it/l} \frac{r}{\sqrt{1 + \frac{r^2}{l^2}}}\,, \label{linecircle} \\
r &\to & l e^{-it/l} \sqrt{1 + \frac{r^2}{l^2}}\,, \nonumber
\end{eqnarray}

maps the AdS$_2^0$ metric (the Ramond ground state) to the AdS$_2^-$ metric
with $a^2 = -1$, i.~e.~, to the Neveu-Schwarz ground state. But near the
boundary $r \to \infty$, Eq.~(\ref{linecircle}) becomes precisely the map
from the line to the circle with circumference $2\pi l$.

\section{Reduction to superconformal quantum mechanics}
\label{redsupconf}

As we mentioned above, the functions $\alpha(t)$, $\beta(t)$,
$\gamma(t)$, $\mu(t)$ and $\nu(t)$ that appear in $\phi$ (\ref{asymptA})
and $\lambda$ (\ref{asymptlambda}) are not arbitrary; imposing the
equations of motion (\ref{DS=0}) one obtains

\begin{eqnarray}
&& T_B'\alpha + 2T_B\alpha' - \frac{c}{12}l^2\alpha''' - \frac l2 T_F'\mu
- \frac 32 l T_F\mu' = 0\,, \label{EqTB} \\
&& T_F'\alpha + \frac 32 T_F\alpha' + 2T_B\mu - \frac c3 l^2\mu'' = 0\,,
\label{EqTF}
\end{eqnarray}

and

\begin{eqnarray}
\frac c6 \beta &=& -T_B\alpha + \frac l2 T_F\mu + \frac{c}{12}l^2\alpha''\,,
\nonumber \\
\frac c3 l\nu &=& T_F\alpha - \frac c3 l^2\mu'\,, \nonumber \\
\gamma &=& -\alpha'\,.
\label{Eqrest}
\end{eqnarray}

For given stress tensor $T_B(t)$ and spin $3/2$ current $T_F(t)$, the
equations (\ref{EqTB}), (\ref{EqTF}) determine $\alpha(t)$ and $\mu(t)$.
From the remaining relations (\ref{Eqrest}) one obtains then the functions
$\beta(t)$, $\gamma(t)$ and $\nu(t)$. Comparing (\ref{EqTB}) and (\ref{EqTF})
with (\ref{transfTB}) and (\ref{transfTF}), we see that $\alpha(t)$ and
$\mu(t)$ correspond to superconformal transformations that leave the stress tensor
and its superpartner invariant. Since Eqns.~(\ref{EqTB}), (\ref{EqTF}) result
from the equation of motion (\ref{DS=0}) for the scalar multiplet $S$,
this is not surprising:
$S$ generates gauge transformations (\ref{gaugetransfGam}) that do not change
the superconnection $\Gamma$.

What is less obvious, and comes as a surprise, is that the dynamics
described by (\ref{EqTB}) and (\ref{EqTF}) is precisely that of
superconformal mechanics. To show this, we multiply Eq.~(\ref{EqTB}) by
$-12\alpha/cl^2$, and add this to (\ref{EqTF}), multiplied (from the right)
by $12\mu/cl$, to obtain

\begin{equation}
\left(\alpha \alpha'' - \frac 12 \alpha'^2 - \frac{12}{cl^2}T_B\alpha^2
+ \frac{18}{cl}T_F\alpha \mu - 4l\mu' \mu\right)' = 0\,. \label{alal''}
\end{equation}

Defining

\begin{equation}
\alpha =: q^2\,,
\end{equation}

one gets from integration of (\ref{alal''})

\begin{equation}
q'' - \frac{g}{q^3} - \frac{6}{cl^2}T_B q - \frac{2l\mu' \mu}{q^3} +
\frac{9}{clq}T_F \mu = 0\,, \label{integrated}
\end{equation}

where $g$ denotes an integration constant. Next, we multiply Eq.~(\ref{EqTF})
by $q$, and add this to (\ref{integrated}), multiplied by $cl^2\mu/3$.
This yields

\begin{equation}
\varsigma^1 = -\frac{q^2}{\sqrt g} {\varsigma^2}'\,,
\label{varsigma1}
\end{equation}

where the spinors $\varsigma^{1,2}$ are defined by

\begin{eqnarray}
\varsigma^1 &:=& \frac lq \mu\,, \\
\varsigma^2 &:=& \frac{q^2}{\sqrt g} {\varsigma^1}' - \frac{3T_F q^3}{cl\sqrt g}\,.
\label{varsigma2}
\end{eqnarray}

Using the definitions of $\varsigma^{1,2}$ in (\ref{integrated}), one
finally obtains

\begin{equation}
q'' - \frac{g}{q^3} - \frac{6}{cl^2}T_B q + \frac{\sqrt g}{lq^3}[\varsigma^1,
\varsigma^2] + \frac{3}{cl^2}T_F \varsigma^1 = 0\,. \label{eomq}
\end{equation}

(\ref{varsigma1}), (\ref{varsigma2}) and (\ref{eomq}) are precisely
the equations of motion of the osp$(2|2)$ superconformal quantum mechanics
considered in \cite{Akulov:uh,Fubini:1984hf} (the ${\cal N} = 2$ superextension
of the DFF model \cite{deAlfaro:1976je}), with an additional coupling of the
fields $q(t)$ and $\varsigma^1(t)$ to the external sources $T_B(t)$ and $T_F(t)$.
These equations of motion result from the action

\begin{equation}
I = \int dt \left(\frac 12 q'^2 - \frac{1}{2l}\varsigma^1{\varsigma^1}'
    - \frac{1}{2l}\varsigma^2{\varsigma^2}' - \frac{g}{2q^2} + \frac{\sqrt g}{2lq^2}
    [\varsigma^1, \varsigma^2] + \frac{3}{cl^2}T_B q^2 - \frac{3}{cl^2}
    T_F \varsigma^1 q\right)\,.
    \label{actionconfqm}
\end{equation}

Note that $T_{B,F}$, being external sources, are not varied.
The partition function computed with the action (\ref{actionconfqm}) is the
generating functional for correlators of the composite operators
$q^2$ and $\varsigma^1 q$.

If we define the ${\cal N} = 1$ superfield

\begin{equation}
\Phi(t) = q(t) + \theta\varsigma^1(t)\,,
\end{equation}

and the supercurrent

\begin{equation}
{\cal T}(t) = T_F(t) + 2\theta T_B(t)\,,
\end{equation}

the source term in (\ref{actionconfqm}) can be written as

\begin{equation}
I_s = \frac{3}{2cl^2}\int dt d\theta\, \Phi^2{\cal T}\,.
\end{equation}

In the bosonic case, the emergence of the DFF model of conformal
quantum mechanics, coupled to an external source, was first shown
in \cite{Cadoni:2000gm}.

It is clear that a model of conformal quantum mechanics like the DFF
model cannot be invariant under the whole Virasoro algebra.
However, when it is coupled to an external field of conformal dimension
2 (e.~g.~, a stress tensor) that transforms appropriately, it can have
such an invariance. The key point is the following observation:\footnote{For
simplicity we consider the bosonic case only. The calculation is easily
generalized to the supersymmetric model.}
Start from the DFF model with action

\begin{equation}
I = \int dt \left(\frac{q'^2}{2} - \frac{g}{2q^2}\right)\,.
    \label{DFFaction}
\end{equation}

If we generalize the transformations considered by DFF \cite{deAlfaro:1976je},
$t = g(\tau)$, $q = Q \sqrt{\dot{g}(\tau)}$, where a dot denotes a derivative
with respect to $\tau$, to arbitrary functions $g(\tau)$, (\ref{DFFaction}) goes over
(modulo total derivatives) into

\begin{equation}
I = \int d\tau \left(\frac{\dot{Q}^2}{2} - \frac{g}{2Q^2} -
    \frac{Q^2}{4}\{g,\tau\}\right)\,, \label{transformedDFF}
\end{equation}

where

\begin{equation}
\{g, \tau\} = \frac{2\dddot g \dot g - 3\ddot{g}^2}{2\dot{g}^2}
\end{equation}

is the Schwarzian derivative. DFF considered only PSL$(2,\bR)$ transformations
with

\begin{equation}
g(\tau) = \frac{A\tau + B}{C\tau + D}\,, \qquad AD - BC = 1\,,
\end{equation}

for which the Schwarzian vanishes, so that the action (\ref{DFFaction}) remains
invariant. One can obtain invariance under the whole Virasoro algebra,
corresponding to arbitrary functions $g(\tau)$, by coupling the field $q$
to an external source $T_B(t)$, that transforms like a stress tensor,

\begin{equation}
\left(\frac{dt}{d\tau}\right)^2 T_B(t) = \tilde{T}_B(\tau) - \frac{c}{12}\{t, \tau\}\,.
\end{equation}

The transformation law of the coupling term $\sim q^2 T_B(t)$ involves then a
Schwarzian derivative that can compensate the one in (\ref{transformedDFF}),
so that the action enjoys invariance under the whole Virasoro algebra.

\section{De~Sitter gravity}
\label{dS}

In this section we generalize the above considerations to the case of
two-dimensional de~Sitter gravity. For related work,
cf.~\cite{Ness:2002qr,Cadoni:2002kz}. Unlike in higher dimensions,
the two-dimensional de~Sitter algebra so$(2,1)$ admits superextensions
that do have unitary highest weight representations. Nevertheless, to keep
things simple, we will consider below the bosonic case only. The inclusion of
supersymmetry should be straightforward.

The action is given by the bosonic part of (\ref{actiongrav}), where now
$\Lambda = -1/l^2$. Adapting the conventions as in appendix \ref{convdS},
this model can again be formulated as an so$(2,1)$ $\approx$ sl$(2,\bR)$
topological gauge theory \cite{Chamseddine:1990wn}, with action

\begin{equation}
I = -\frac{\beta}{2} \int {\mathrm{Tr}}(\phi F)\,,
\end{equation}

where $F = dA + A \wedge A$ denotes the sl$(2,\bR)$ field strength, and $\phi$
is an sl$(2,\bR)$ valued scalar field.
The gravity action is recovered by decomposing the connection $A$ and the
scalar $\phi$ according to (\ref{decomp}).

Up to trivial gauge transformations, the boundary conditions for
asymptotically ($t \to \infty$) de~Sitter spaces given in \cite{Brigante:2002rv}
read in terms of the gauge fields

\begin{eqnarray}
A &=& \left[\frac{t}{l^2} - \frac{T_B(\sigma)}{2\eta_0 t}\right] \tau_1 d\sigma
        + \left[\frac{t}{l^2} + \frac{T_B(\sigma)}{2\eta_0 t}\right] \tau_2 d\sigma
        + \frac{dt}{t} \tau_0\,, \nonumber \\
\phi &=& \left[\alpha(\sigma)\frac{t}{l^2} - \frac{\beta(\sigma)}{t}\right] \tau_1
         + \left[\alpha(\sigma)\frac{t}{l^2} + \frac{\beta(\sigma)}{t}\right] \tau_2
         + \gamma(\sigma) \tau_0\,, \label{dSasympt}
\end{eqnarray}

where the functions $T_B(\sigma)$, $\alpha(\sigma)$, $\beta(\sigma)$ and
$\gamma(\sigma)$, that depend on the boundary coordinate $\sigma$,
parametrize the subleading behaviour. The meaning of the constant $\eta_0$ will
become clear below.
Notice that the connection $A$ in (\ref{dSasympt}) is flat for arbitrary
$T_B(\sigma)$. The metric

\begin{equation}
ds^2 = \frac{l^2}{t^2}dt^2 - \left(\frac tl - \frac{T_B(\sigma)l}{2\eta_0 t}\right)^2
       d\sigma^2\,, \label{genmetrdS}
\end{equation}

resulting from (\ref{dSasympt}) satisfies thus
the field equations for any $T_B(\sigma)$.

Let us now consider the solutions

\begin{eqnarray}
ds^2 &=& \left(\frac{t^2}{l^2} - a^2\right)^{-1}dt^2 -
         \left(\frac{t^2}{l^2} - a^2\right) d\sigma^2\,, \label{dScad} \\
\eta &=& \eta_0 \frac tl\,, \label{etadS}
\end{eqnarray}

studied in \cite{Cadoni:2002kz}, with $a$ and $\eta_0 \ge 0$ denoting
integration constants. In the static patch $0 \le t \le al$, $t$ becomes
a spacelike coordinate, whereas $\sigma$ becomes timelike. One
can associate to the cosmological event horizon at $t = al$ the mass
and entropy \cite{Cadoni:2002kz}

\begin{equation}
M = \frac{1}{2l}\eta_0 a^2\,, \qquad S = 2\pi \eta_0 a\,.
    \label{MSdS}
\end{equation}

Applying the diffeomorphism $t \mapsto t + a^2l^2/4t$ to the metric
(\ref{dScad}) yields

\begin{equation}
ds^2 =  \frac{l^2}{t^2} dt^2 - \left(\frac tl - \frac{a^2l}{4t}\right)^2 d\sigma^2\,.
\end{equation}

Comparing this with (\ref{genmetrdS}), we see that in this case one has
$T_B(t) = \eta_0 a^2/2 = Ml$, so it essentially coincides with the de~Sitter
mass.

The gauge transformations

\begin{eqnarray}
\delta_{\Lambda} F &=& dF + [F, \Lambda]\,, \nonumber \\
\delta_{\Lambda} \phi &=& [\phi, \Lambda]\,,
\end{eqnarray}

that preserve the asymptotic behaviour (\ref{dSasympt}) are generated by

\begin{eqnarray}
\Lambda^0 &=& -\chi'(\sigma)\,, \nonumber \\
\Lambda^1 &=& \left(\frac{t}{l^2} - \frac{T_B(\sigma)}{2\eta_0 t}\right)\chi(\sigma)
              + \frac{l^2}{2t}\chi''(\sigma)\,, \label{pardS} \\
\Lambda^2 &=& \left(\frac{t}{l^2} + \frac{T_B(\sigma)}{2\eta_0 t}\right)\chi(\sigma)
              - \frac{l^2}{2t}\chi''(\sigma)\,. \nonumber
\end{eqnarray}

As the function $\chi(\sigma)$ is arbitrary, the asymptotic symmetry algebra
is infinite-dimensional. Under a gauge transformation with parameters
(\ref{pardS}), $T_B(\sigma)$ transforms like a stress tensor,

\begin{equation}
\delta T_B = T_B'\chi + 2T_B\chi' - \frac{c}{12}l^2\chi'''\,,
\end{equation}

with central charge

\begin{equation}
c = 12\eta_0\,.
\end{equation}

This central charge appearing in the algebra of asymptotic symmetries was
found by different methods in \cite{Cadoni:2002kz}.

Using the Cardy formula (\ref{cardy}) with $L_0 = Ml$, and mass $M$ given
by Eq.~(\ref{MSdS}), we recover precisely the thermodynamical entropy of
de~Sitter space.

On shell, one finds for the functions $\alpha(\sigma)$, $\beta(\sigma)$ and
$\gamma(\sigma)$ that parametrize the subleading behaviour of the scalar
field $\phi$,

\begin{eqnarray}
T_B'\alpha &+& 2T_B\alpha' - \frac{c}{12}l^2\alpha''' = 0\,, \label{dynal} \\
\frac c6 \beta &=& T_B\alpha - \frac{c}{12}l^2\alpha''\,, \nonumber \\
\gamma &=& -\alpha'\,. \nonumber
\end{eqnarray}

Again, the dynamics (\ref{dynal}) of the field $\alpha$ is that of conformal
transformations that leave the CFT stress tensor $T_B$ invariant. Like in
section \ref{redsupconf}, defining $\alpha =: q^2$, one finds that
(\ref{dynal}) is equivalent to the equation of motion following from the
action

\begin{equation}
I = \int d\sigma \left(\frac 12 q'^2 - \frac{g}{2q^2} + \frac{3}{cl^2}T_B q^2\right)\,,
\end{equation}

which is the DFF model\footnote{Aspects of the DFF model in the context of
particle motion in spacetimes containing Killing horizons are studied
in \cite{Moretti:2002mp}.} coupled to the spin two current $T_B$.

Let us finally comment on the fact that dS$_2$ has two conformal boundaries,
the past boundary ${\cal I}^-$ and the future boundary ${\cal I}^+$,
so in principle one expects a dual description in terms of a quantum
mechanical system living on a disconnected manifold. Of course the same
problem appears in the AdS$_2$ case, since two-dimensional anti-de~Sitter space,
unlike its higher-dimensional cousins, has two boundaries. A possible way
to eliminate the second boundary is to cut the spacetime at
$r=0$ ($t=0$ in the dS$_2$ case),
in order that the position-dependent coupling constant represented
by the dilaton (cf.~Eqns.~(\ref{eta}) and (\ref{etadS})) is
positive \cite{Cadoni:1999ja}. Note in this context that the coordinates used
e.~g.~in (\ref{dScad}) are not global for $a^2 > 0$, so using this
coordinate system we implicitely restricted ourselves to one boundary
only.

\section{Final remarks}

In this paper we investigated the asymptotic dynamics of topological
anti-de~Sitter supergravity in two dimensions.
We showed that the asymptotic symmetries
form a super-Virasoro algebra. Using the central charge of this algebra in
Cardy's formula, we exactly reproduced the thermodynamical entropy of AdS$_2$
black holes. Furthermore, we saw that the dynamics of the dilaton and its
superpartner reduces to that of superconformal transformations that leave
invariant one holomorphic component of the stress tensor supercurrent of a
two-dimensional conformal field theory. It was then shown that this dynamics
is governed by a supersymmetric extension of the de~Alfaro-Fubini-Furlan model
of conformal quantum mechanics.

We also considered two-dimensional de~Sitter gravity, and computed the dS$_2$
entropy by counting CFT states. Again, the result coincides with the
thermodynamical entropy of dS$_2$.

We do not expect that the superconformal structure discovered here
is due to the topological nature of the underlying supergravity model.
This can be checked by considering more general (non topological) supergravity
theories\footnote{Note in this context that also the CGHS model of two-dimensional
dilaton gravity can be formulated as a topological gauge theory, based on the
centrally extended Poincar\'{e} group \cite{Cangemi:1992bj}.}
that admit AdS$_2$ vacua, studying the asymptotic symmetries and
determining the dynamics of the fields that parametrize the subleading
asymptotic behaviour. Such an analysis should be considerably simplified using
the formulation of two-dimensional dilaton gravity as a
Poisson-Sigma-model \cite{Klosch:1995fi}\footnote{See \cite{Grumiller:2002nm} for a
review of dilaton gravity in two dimensions.}.
Work in this direction is in progress.

\section*{Acknowledgements}
\small

This work was partially supported by INFN, MURST and
by the European Commission RTN program
HPRN-CT-2000-00131, in which S.~C.~, D.~K.~and D.~Z.~are
associated to the University of Torino.
\normalsize

\newpage

\begin{appendix}

\section{Conventions}

\subsection{Anti-de~Sitter}
\label{convAdS}

We denote the bosonic osp$(1|2)$ generators by $\tau_A$, $A = 0,1,2$, and
the fermionic ones by $R^i$, $i=1,2$. The $\tau_A$ satisfy the sl$(2,\bR)$
commutation relations

\begin{equation}
[\tau_A, \tau_B] = -\epsilon_{ABC}\tau^C\,,
\end{equation}

where $\epsilon_{012} = 1$, and indices are raised and lowered with
the metric $(\eta_{AB}) = {\mathrm{diag}}(1,-1,-1)$.
Furthermore one has

\begin{equation}
[\tau_A, R^i] = f_{A\;\;j}^{\;\;i} R^j\,, \qquad \{R^i, R^j\} = f^{ij}_{\;\;A}
\tau^A\,,
\end{equation}

with $f_{A\;\;j}^{\;\;i} = -(\tau_A)^i_{\;\,j}$ and
$f^{ij}_{\;\;A} = \epsilon^{ik}(\tau_A)^j_{\;\,k}$, $\epsilon^{12} = 1$.
We choose the explicit representation

\begin{equation}
\tau_0 = \frac 12\left(\begin{array}{cc} 0 & 1 \\ -1 & 0 \end{array}
                 \right)\,, \quad
\tau_1 = -\frac 12\left(\begin{array}{cc} 0 & 1 \\ 1 & 0 \end{array}
                 \right)\,, \quad
\tau_2 = \frac 12\left(\begin{array}{cc} -1 & 0 \\ 0 & 1 \end{array}
                 \right)\, \label{generatorsAdS}
\end{equation}

of the sl$(2,\bR)$ generators.

The superalgebra osp$(1|2)$ admits an invariant, supersymmetric, nondegenerate
bilinear form ${\mathrm{STr}}$ defined by

\begin{equation}
{\mathrm{STr}}(\tau_A\tau_B) = -\frac 12 \eta_{AB}\,, \qquad
{\mathrm{STr}}(R^i R^j) = -\frac 12 \epsilon^{ij}\,.
\end{equation}

For the gamma matrices in two dimensions we take $\gamma^a = 2i\tau^a$, where
$a = 0,1$. They satisfy

\begin{equation}
\{\gamma^a, \gamma^b\} = 2\eta^{ab}\,,
\end{equation}

with $(\eta^{ab}) = {\mathrm{diag}}(1,-1)$. Furthermore we defined
$\gamma^{ab} = \frac 12 [\gamma^a, \gamma^b]$, and the parity matrix
$\gamma_5 = \gamma_0 \gamma_1$.

Finally, $dt \wedge dr$ is chosen to have positive orientation.

\subsection{De~Sitter}
\label{convdS}

In order to cover the case of de~Sitter gravity, we choose the
sl$(2,\bR)$ generators $\tau_A$ such that

\begin{equation}
{\mathrm{Tr}}(\tau_A\tau_B) = \frac 12 \eta_{AB}\,,
\end{equation}

where now $(\eta_{AB}) = {\mathrm{diag}}(1,-1,1)$. A possible representation is

\begin{equation}
\tau_0 = \frac 12\left(\begin{array}{cc} 1 & 0 \\ 0 & -1 \end{array}
                 \right)\,, \quad
\tau_1 = \frac 12\left(\begin{array}{cc} 0 & 1 \\ -1 & 0 \end{array}
                 \right)\,, \quad
\tau_2 = -\frac 12\left(\begin{array}{cc} 0 & 1 \\ 1 & 0 \end{array}
                 \right)\,.
\end{equation}

\end{appendix}


\newpage

\begin{thebibliography}{99}

%\cite{Maldacena:1997re}
\bibitem{Maldacena:1997re}
J.~Maldacena,
``The large $N$ limit of superconformal field theories and supergravity,''
Adv.\ Theor.\ Math.\ Phys.\  {\bf 2} (1998) 231
[Int.\ J.\ Theor.\ Phys.\  {\bf 38} (1998) 1113]
[arXiv:hep-th/9711200].
%%CITATION = HEP-TH 9711200;%%

%\cite{Witten:1998qj}
\bibitem{Witten:1998qj}
E.~Witten,
``Anti-de Sitter space and holography,''
Adv.\ Theor.\ Math.\ Phys.\  {\bf 2} (1998) 253
[arXiv:hep-th/9802150].
%%CITATION = HEP-TH 9802150;%%

%\cite{'tHooft:gx}
\bibitem{'tHooft:gx}
G.~'t Hooft,
``Dimensional Reduction In Quantum Gravity,''
arXiv:gr-qc/9310026;\\
%%CITATION = GR-QC 9310026;%%
%\cite{Susskind:1994vu}
L.~Susskind,
``The World as a hologram,''
J.\ Math.\ Phys.\  {\bf 36} (1995) 6377
[arXiv:hep-th/9409089].
%%CITATION = HEP-TH 9409089;%%

%\cite{Coussaert:1995zp}
\bibitem{Coussaert:1995zp}
O.~Coussaert, M.~Henneaux and P.~van Driel,
``The asymptotic dynamics of three-dimensional Einstein gravity with a
negative cosmological constant,''
Class.\ Quant.\ Grav.\  {\bf 12} (1995) 2961
[arXiv:gr-qc/9506019].
%%CITATION = GR-QC 9506019;%%

%\cite{Achucarro:1986vz}
\bibitem{Achucarro:1986vz}
A.~Achucarro and P.~K.~Townsend,
``A Chern-Simons Action For Three-Dimensional Anti-De Sitter
Supergravity Theories,''
Phys.\ Lett.\ B {\bf 180} (1986) 89;
%%CITATION = PHLTA,B180,89;%%

%\cite{Witten:1988hc}
\bibitem{Witten:1988hc}
E.~Witten,
``(2+1)-Dimensional Gravity As An Exactly Soluble System,''
Nucl.\ Phys.\ B {\bf 311} (1988) 46.
%%CITATION = NUPHA,B311,46;%%

%\cite{Elitzur:1989nr}
\bibitem{Elitzur:1989nr}
S.~Elitzur, G.~W.~Moore, A.~Schwimmer and N.~Seiberg,
``Remarks On The Canonical Quantization Of The Chern-Simons-Witten Theory,''
Nucl.\ Phys.\ B {\bf 326} (1989) 108.
%%CITATION = NUPHA,B326,108;%%

%\cite{Brown:nw}
\bibitem{Brown:nw}
J.~D.~Brown and M.~Henneaux,
``Central Charges In The Canonical Realization Of Asymptotic Symmetries:
An Example From Three-Dimensional Gravity,''
Commun.\ Math.\ Phys.\  {\bf 104} (1986) 207.
%%CITATION = CMPHA,104,207;%%

%\cite{Henneaux:1999ib}
\bibitem{Henneaux:1999ib}
M.~Henneaux, L.~Maoz and A.~Schwimmer,
``Asymptotic dynamics and asymptotic symmetries of three-dimensional
extended AdS supergravity,''
Annals Phys.\  {\bf 282} (2000) 31
[arXiv:hep-th/9910013].
%%CITATION = HEP-TH 9910013;%%

%\cite{Cacciatori:2001un}
\bibitem{Cacciatori:2001un}
S.~Cacciatori and D.~Klemm,
``The asymptotic dynamics of de Sitter gravity in three dimensions,''
Class.\ Quant.\ Grav.\  {\bf 19} (2002) 579
[arXiv:hep-th/0110031].
%%CITATION = HEP-TH 0110031;%%

%\cite{Klemm:2002ir}
\bibitem{Klemm:2002ir}
D.~Klemm and L.~Vanzo,
``De~Sitter gravity and Liouville theory,''
JHEP {\bf 0204} (2002) 030
[arXiv:hep-th/0203268].
%%CITATION = HEP-TH 0203268;%%

%\cite{Hull:1998vg}
\bibitem{Hull:1998vg}
C.~M.~Hull,
``Timelike T-duality, de~Sitter space, large $N$ gauge theories and
topological field theory,''
JHEP {\bf 9807} (1998) 021
[arXiv:hep-th/9806146].
%%CITATION = HEP-TH 9806146;%%

%\cite{Strominger:2001pn}
\bibitem{Strominger:2001pn}
A.~Strominger,
``The dS/CFT correspondence,''
JHEP {\bf 0110} (2001) 034
[arXiv:hep-th/0106113].
%%CITATION = HEP-TH 0106113;%%

%\cite{Klemm:2001ea}
\bibitem{Klemm:2001ea}
D.~Klemm,
``Some aspects of the de Sitter/CFT correspondence,''
Nucl.\ Phys.\ B {\bf 625} (2002) 295
[arXiv:hep-th/0106247].
%%CITATION = HEP-TH 0106247;%%

%\cite{Park:1998qk}
\bibitem{Park:1998qk}
M.~I.~Park,
``Statistical entropy of three-dimensional Kerr-de Sitter space,''
Phys.\ Lett.\ B {\bf 440} (1998) 275
[arXiv:hep-th/9806119].
%%CITATION = HEP-TH 9806119;%%

%\cite{Strominger:1998yg}
\bibitem{Strominger:1998yg}
A.~Strominger,
``AdS$_2$ quantum gravity and string theory,''
JHEP {\bf 9901} (1999) 007
[arXiv:hep-th/9809027].
%%CITATION = HEP-TH 9809027;%%

%\cite{Claus:1998ts}
\bibitem{Claus:1998ts}
P.~Claus, M.~Derix, R.~Kallosh, J.~Kumar, P.~K.~Townsend and A.~Van Proeyen,
``Black holes and superconformal mechanics,''
Phys.\ Rev.\ Lett.\  {\bf 81} (1998) 4553
[arXiv:hep-th/9804177].
%%CITATION = HEP-TH 9804177;%%

%\cite{Cacciatori:1999rp}
\bibitem{Cacciatori:1999rp}
S.~Cacciatori, D.~Klemm and D.~Zanon,
``$w_{\infty}$ algebras, conformal mechanics, and black holes,''
Class.\ Quant.\ Grav.\  {\bf 17} (2000) 1731
[arXiv:hep-th/9910065].
%%CITATION = HEP-TH 9910065;%%

%\cite{Cadoni:2000gm}
\bibitem{Cadoni:2000gm}
M.~Cadoni, P.~Carta, D.~Klemm and S.~Mignemi,
``AdS$_2$ gravity as conformally invariant mechanical system,''
Phys.\ Rev.\ D {\bf 63} (2001) 125021
[arXiv:hep-th/0009185].
%%CITATION = HEP-TH 0009185;%%

%\cite{Teitelboim:ux}
\bibitem{Teitelboim:ux}
C.~Teitelboim,
``Gravitation And Hamiltonian Structure In Two Space-Time Dimensions,''
Phys.\ Lett.\ B {\bf 126} (1983) 41, and ``The Hamiltonian structure of
two-dimensional space-time and its relation with the conformal anomaly,''
in {\it Quantum Theory of Gravity},
S.~Christensen, ed.~(Adam Hilger, Bristol, 1984);\\
%%CITATION = PHLTA,B126,41;%%
%\cite{Jackiw:1984je}
R.~Jackiw, ``Liouville field theory: a two-dimensional model for gravity?,''
in {\it Quantum Theory of Gravity},
S.~Christensen, ed.~(Adam Hilger, Bristol, 1984), and
``Lower Dimensional Gravity,''
Nucl.\ Phys.\ B {\bf 252} (1985) 343.
%%CITATION = NUPHA,B252,343;%%

%\cite{deAlfaro:1976je}
\bibitem{deAlfaro:1976je}
V.~de Alfaro, S.~Fubini and G.~Furlan,
``Conformal Invariance In Quantum Mechanics,''
Nuovo Cim.\ A {\bf 34} (1976) 569.
%%CITATION = NUCIA,A34,569;%%

%\cite{Brigante:2002rv}
\bibitem{Brigante:2002rv}
M.~Brigante, S.~Cacciatori, D.~Klemm and D.~Zanon,
``The asymptotic dynamics of two-dimensional (anti-)de Sitter gravity,''
JHEP {\bf 0203} (2002) 005
[arXiv:hep-th/0202073].
%%CITATION = HEP-TH 0202073;%%

%\cite{Isler:1989hq}
\bibitem{Isler:1989hq}
K.~Isler and C.~A.~Trugenberger,
``A Gauge Theory Of Two-Dimensional Quantum Gravity,''
Phys.\ Rev.\ Lett.\  {\bf 63} (1989) 834.
%%CITATION = PRLTA,63,834;%%

%\cite{Chamseddine:1990wn}
\bibitem{Chamseddine:1990wn}
A.~H.~Chamseddine and D.~Wyler,
``Topological Gravity In (1+1)-Dimensions,''
Nucl.\ Phys.\ B {\bf 340} (1990) 595.
%%CITATION = NUPHA,B340,595;%%

%\cite{Cadoni:1998sg}
\bibitem{Cadoni:1998sg}
M.~Cadoni and S.~Mignemi,
``Entropy of 2D black holes from counting microstates,''
Phys.\ Rev.\ D {\bf 59} (1999) 081501
[arXiv:hep-th/9810251].
%%CITATION = HEP-TH 9810251;%%

%\cite{Cadoni:1999ja}
\bibitem{Cadoni:1999ja}
M.~Cadoni and S.~Mignemi,
``Asymptotic symmetries of AdS$_2$ and conformal group in d = 1,''
Nucl.\ Phys.\ B {\bf 557} (1999) 165
[arXiv:hep-th/9902040].
%%CITATION = HEP-TH 9902040;%%

%\cite{Chamseddine:1991fg}
\bibitem{Chamseddine:1991fg}
A.~H.~Chamseddine,
``Superstrings in arbitrary dimensions,''
Phys.\ Lett.\ B {\bf 258} (1991) 97.
%%CITATION = PHLTA,B258,97;%%

%\cite{Schaller:1993pa}
\bibitem{Schaller:1993pa}
P.~Schaller and T.~Strobl,
``Diffeomorphisms versus nonabelian gauge transformations: An example of
(1+1)-dimensional gravity,''
Phys.\ Lett.\ B {\bf 337} (1994) 266
[arXiv:hep-th/9401110].
%%CITATION = HEP-TH 9401110;%%

%\cite{Banados:1998gg}
\bibitem{Banados:1998gg}
M.~Ba\~{n}ados,
``Three-dimensional quantum geometry and black holes,''
arXiv:hep-th/9901148.
%%CITATION = HEP-TH 9901148;%%

%\cite{Christensen:1991dk}
\bibitem{Christensen:1991dk}
D.~Christensen and R.~B.~Mann,
``The Causal structure of two-dimensional space-times,''
Class.\ Quant.\ Grav.\  {\bf 9} (1992) 1769
[arXiv:hep-th/9203050];\\
%%CITATION = HEP-TH 9203050;%%
%\cite{Achucarro:1993fd}
A.~Achucarro and M.~E.~Ortiz,
``Relating black holes in two-dimensions and three-dimensions,''
Phys.\ Rev.\ D {\bf 48} (1993) 3600
[arXiv:hep-th/9304068];\\
%%CITATION = HEP-TH 9304068;%%
%\cite{Lemos:1993qn}
J.~P.~Lemos and P.~M.~Sa,
``Nonsingular constant curvature two-dimensional black hole,''
Mod.\ Phys.\ Lett.\ A {\bf 9} (1994) 771
[arXiv:gr-qc/9309023];
%%CITATION = GR-QC 9309023;%%
%\cite{Lemos:py}
``The Black Holes Of A General Two-Dimensional Dilaton Gravity Theory,''
Phys.\ Rev.\ D {\bf 49} (1994) 2897
[Erratum-ibid.\ D {\bf 51} (1994) 5967]
[arXiv:gr-qc/9311008].
%%CITATION = GR-QC 9311008;%%

%\cite{Cadoni:1994uf}
\bibitem{Cadoni:1994uf}
M.~Cadoni and S.~Mignemi,
``Nonsingular four-dimensional black holes and the Jackiw-Teitelboim theory,''
Phys.\ Rev.\ D {\bf 51} (1995) 4319
[arXiv:hep-th/9410041].
%%CITATION = HEP-TH 9410041;%%

%\cite{Henneaux:tv}
\bibitem{Henneaux:tv}
M.~Henneaux and C.~Teitelboim,
``Asymptotically Anti-De Sitter Spaces,''
Commun.\ Math.\ Phys.\  {\bf 98} (1985) 391.
%%CITATION = CMPHA,98,391;%%

%\cite{Banados:1998pi}
\bibitem{Banados:1998pi}
M.~Ba\~{n}ados, K.~Bautier, O.~Coussaert, M.~Henneaux and M.~Ortiz,
%``Anti-de Sitter/CFT correspondence in three-dimensional supergravity,''
Phys.\ Rev.\ D {\bf 58} (1998) 085020
[arXiv:hep-th/9805165].
%%CITATION = HEP-TH 9805165;%%

%\cite{Cadoni:2000ah}
\bibitem{Cadoni:2000ah}
M.~Cadoni and S.~Mignemi,
``Symmetry breaking, central charges and the AdS$_2$/CFT$_1$ correspondence,''
Phys.\ Lett.\ B {\bf 490} (2000) 131
[arXiv:hep-th/0002256].
%%CITATION = HEP-TH 0002256;%%

%\cite{Witten:nf}
\bibitem{Witten:nf}
E.~Witten,
``Dynamical Breaking Of Supersymmetry,''
Nucl.\ Phys.\ B {\bf 188} (1981) 513.
%%CITATION = NUPHA,B188,513;%%

%\cite{Caldarelli:2000xk}
\bibitem{Caldarelli:2000xk}
M.~Caldarelli, G.~Catelani and L.~Vanzo,
``Action, Hamiltonian and CFT for 2D black holes,''
JHEP {\bf 0010} (2000) 005
[arXiv:hep-th/0008058].
%%CITATION = HEP-TH 0008058;%%

%\cite{Catelani:2000gn}
\bibitem{Catelani:2000gn}
G.~Catelani and L.~Vanzo,
``On the $\sqrt 2$ puzzle in AdS$_2$/CFT$_1$,''
arXiv:hep-th/0009186.
%%CITATION = HEP-TH 0009186;%%

%\cite{Cardy:ie}
\bibitem{Cardy:ie}
J.~L.~Cardy,
``Operator Content Of Two-Dimensional Conformally Invariant Theories,''
Nucl.\ Phys.\ B {\bf 270} (1986) 186.
%%CITATION = NUPHA,B270,186;%%

%\cite{Coussaert:jp}
\bibitem{Coussaert:jp}
O.~Coussaert and M.~Henneaux,
``Supersymmetry Of The (2+1) Black Holes,''
Phys.\ Rev.\ Lett.\  {\bf 72} (1994) 183
[arXiv:hep-th/9310194].
%%CITATION = HEP-TH 9310194;%%

%\cite{Strominger:1997eq}
\bibitem{Strominger:1997eq}
A.~Strominger,
``Black hole entropy from near-horizon microstates,''
JHEP {\bf 9802} (1998) 009
[arXiv:hep-th/9712251].
%%CITATION = HEP-TH 9712251;%%

%\cite{Akulov:uh}
\bibitem{Akulov:uh}
V.~P.~Akulov and A.~I.~Pashnev,
``Quantum Superconformal Model In (1,2) Space,''
Theor.\ Math.\ Phys.\  {\bf 56} (1983) 862
[Teor.\ Mat.\ Fiz.\  {\bf 56} (1983) 344].
%%CITATION = TMPHA,56,862;%%

%\cite{Fubini:1984hf}
\bibitem{Fubini:1984hf}
S.~Fubini and E.~Rabinovici,
``Superconformal Quantum Mechanics,''
Nucl.\ Phys.\ B {\bf 245} (1984) 17.
%%CITATION = NUPHA,B245,17;%%

%\cite{Ness:2002qr}
\bibitem{Ness:2002qr}
S.~Ness and G.~Siopsis,
``dS/CFT correspondence in two dimensions,''
Phys.\ Lett.\ B {\bf 536} (2002) 315
[arXiv:hep-th/0202096].
%%CITATION = HEP-TH 0202096;%%

%\cite{Cadoni:2002kz}
\bibitem{Cadoni:2002kz}
M.~Cadoni, P.~Carta, M.~Cavagli\`{a} and S.~Mignemi,
``Two-dimensional dS/CFT correspondence,''
Phys.\ Rev.\ D {\bf 66} (2002) 065008
[arXiv:hep-th/0205211].
%%CITATION = HEP-TH 0205211;%%

%\cite{Moretti:2002mp}
\bibitem{Moretti:2002mp}
V.~Moretti and N.~Pinamonti,
``Aspects of hidden and manifest SL$(2,\bR)$ symmetry in 2D near-horizon
black-hole backgrounds,''
arXiv:gr-qc/0207072.
%%CITATION = GR-QC 0207072;%%

%\cite{Cangemi:1992bj}
\bibitem{Cangemi:1992bj}
D.~Cangemi and R.~Jackiw,
``Gauge invariant formulations of lineal gravity,''
Phys.\ Rev.\ Lett.\  {\bf 69} (1992) 233
[arXiv:hep-th/9203056];
%%CITATION = HEP-TH 9203056;%%
%\cite{Cangemi:1994iz}
``Quantum states of string inspired lineal gravity,''
Phys.\ Rev.\ D {\bf 50} (1994) 3913
[arXiv:hep-th/9403045].
%%CITATION = HEP-TH 9403045;%%

%\cite{Klosch:1995fi}
\bibitem{Klosch:1995fi}
T.~Kl\"osch and T.~Strobl,
``Classical and quantum gravity in (1+1)-Dimensions. Part 1: A Unifying approach,''
Class.\ Quant.\ Grav.\  {\bf 13} (1996) 965
[Erratum-ibid.\  {\bf 14} (1997) 825]
[arXiv:gr-qc/9508020].
%%CITATION = GR-QC 9508020;%%

%\cite{Grumiller:2002nm}
\bibitem{Grumiller:2002nm}
D.~Grumiller, W.~Kummer and D.~V.~Vassilevich,
``Dilaton gravity in two dimensions,''
Phys.\ Rept.\  {\bf 369} (2002) 327
[arXiv:hep-th/0204253].
%%CITATION = HEP-TH 0204253;%%

\end{thebibliography}
\end{document}